\documentclass[a4paper,12pt]{article}
\usepackage{amssymb,amsfonts,amsmath,mathtext,cite,enumerate,float}
\usepackage{amssymb,amsthm,latexsym,euscript,amscd, authblk}
\usepackage{braket}

\usepackage[english]{babel}
\usepackage[cp1251]{inputenc}

\newtheorem{definition}{Definition}
\newtheorem{lemma}{Lemma}
\newtheorem{proposition}{Proposition}
\newtheorem{corollary}{Corollary}
\newtheorem{theorem}{Theorem}
\newcommand{\cH}{{\cal H}}
\newcommand{\rank}{{\rm rank}\,}

\title{Non-commutative graphs and quantum error correction for a two-mode quantum oscillator}

\author[1]{G.G. Amosov\thanks {gramos@mi-ras.ru}}

\author[1]{A.S. Mokeev\thanks {alexandrmokeev@yandex.ru}}

\author[1]{A.N. Pechen\thanks {apechen@gmail.com}}

\affil[1] {Steklov Mathematical Institute of Russian Academy of Sciences, Gubkina str., 8, Moscow 119991, Russia}

\begin{document}

\maketitle

\begin{abstract}
An important topic in quantum information is the theory of error correction codes. Practical situations often involve quantum systems with states in an infinite dimensional Hilbert space, for example coherent states. Motivated by these practical needs, we develop the theory of non-commutative graphs, which is a tool to analyze error correction codes, to infinite dimensional Hilbert spaces. As an explicit example, a family of non-commutative graphs associated with the Schr\"odinger equation describing the dynamics of a two-mode quantum oscillator is constructed and  maximal quantum anticliques for these graphs are found. 
\end{abstract}


\section{Introduction}

An important topic in quantum information is the theory of error correction codes~\cite{HolevoBook,Shor1995,Knill1997,Knill2000,Gaitan2008,Lidar2013} used to protect quantum information from decoherence and other noises.  The information is encoded in the density matrix $\rho$ of the system. Given a quantum channel $\Phi$ (i.e., a completely positive trace preserving map) which describes information transmission from sender to receiver and includes noise and other destructive for information transmission effects, an error correction code is a set of states in $\cal H$ which can be exactly distinguished  after a transmission via the channel $\Phi$. Quantum error correcting codes are actively studied theoretically~\cite{Gottesman2001,Kribs1, Kosut2009,Kribs3, Kribs4, Albert2018} and experimentally~\cite{Schindler2011,Reed2012}. They can rely on encoding the information in finite-dimensional states of a quantum system, as was originally considered for example in~\cite{Shor1995}, or in infinite-dimensional states as for example using states of a quantum oscillator~\cite{Gottesman2001,Albert2018}. Continuous variables are also used, e.g, for generating continuous variable quadripartite cluster and Greenberger-Horne-Zeilinger entangled states of electromagnetic fields~\cite {Su2007}, studying attacks and security analysis for protocols of quantum cryptography~\cite{Kronberg2018,Bochkov2019}, etc. Controlling a two-mode quantum oscillator already appears a nontrivial complex NP-complete problem~\cite{Bondar2019}.

Every quantum channel has a Kraus decomposition $\Phi (\rho )=\sum_jV_j\rho V_j^*$ with some operators $V_j$. The crucial role in finding an appropriate error correction code is played by the linear space $\mathcal V$ spanned by the operators $V_j^*V_k$. The famous Knill--Laflamme condition claims that a zero error transmission via channel $\Phi $ is possible iff $P{\mathcal V}P={\mathbb C}P$ for the orthogonal projection $P$ on the subspace generated by an error correction code~\cite{Knill1997}. Throughout this paper we call $\mathcal V$ a non-commutative graph.
 
In~\cite{amo, amosovmokeev2, amosovmokeev3} the study of non-commutative graphs generated by covariant resolutions of identity was initiated for finite-dimensional Hilbert spaces. The cases of the commutative circle group~\cite{amosovmokeev2} and the non-commutative Heisenberg--Weyl group~\cite{amosovmokeev3}
were considered. In this work we consider non-commutative graphs for an infinite dimensional space $\mathcal H$. An example of a non-commutative operator graph generated by the dynamics of diatomic molecule, or coupled oscillators, is explicitly constructed and its maximal anticlique is found.

The paper is organized as follows. Section~\ref{Prelim} is devoted to the introduction of basic notions of the theory
of non-commutative graphs for analyzing error correction
codes in infinite dimensional Hilbert spaces. In Section~\ref{Sec:Explicit} the action of the unitary group used for a generation of graph is obtained in the explicit form. In Section~\ref{Sec:Graph} a family of graphs generated by this group is constructed and in Section~\ref{Sec:Maximal} their maximal quantum anticliques are found. In the Conclusion Section we give some conclusive remarks.

\section{Non-commutative graphs generated by orbits of unitary groups}\label{Prelim}

\begin{definition} A {\bf non-commutative graph} is a linear subspace $\mathcal V$ of bounded
operators in a Hilbert space $\cH$ possessing the properties
\begin{itemize}
\item ${\bf V}\in \mathcal V$ implies that ${\bf V}^*\in \mathcal V$;
\item ${\bf I}\in {\mathcal V}$
\end{itemize}
\end{definition}
Such objects were introduced in~\cite{ChoiEffros} as operator systems and recently redefined in quantum information theory under the name of
a non-commutative graph~\cite{Duan}. 

In finite-dimensional space, an operator graph $\mathcal V$ can be generated by a unitary representation $g\to {\bf U}_g$ in a Hilbert space $\cH$ of a compact group $G$ with the Haar measure $\mu$ as
\begin{equation}\label{graph}
{\mathcal V}=span\{{\bf U}_g{\bf Q}{\bf U}_g^*,\ g\in G\},
\end{equation}
Here $\bf Q$ is an orthogonal projection such that
\begin{equation}\label{proj}
\int \limits _{G}{\bf U}_g{\bf Q}{\bf U}_g^*d\mu (g)={\bf I}
\end{equation}
to guarantee that ${\bf I}\in {\mathcal V}$. At the moment, only finite dimensional case $\dim\cH<+\infty$ has been studied in this framework. 
Below we consider the infinite dimensional case $\dim {\cal H}=+\infty$. 

Suppose that the dynamics of the quantum system is determined by the Schr\"odinger equation
\begin{equation}\label{sch}
i\psi _t={\bf H}\psi .
\end{equation} 
Then instead of~(\ref {graph}) and~(\ref {proj}),
we shall search for a linear subspace $\mathcal V$ and a set of orthogonal projections ${\bf Q}_{\beta }$ with a parameter $\beta $ belonging to some set $\mathfrak {B}$ such that
\begin{equation}\label{uslovie}
{\mathcal V}=\overline {span\{{\bf U}_t{\bf Q}_{\beta }{\bf U}_t^*,\ t\in{\mathbb R},\ \beta \in \mathfrak {B}\}},
\end{equation}
\begin{equation}\label{uslovie2}
{\bf I}\in {\mathcal V},
\end{equation}
where $({\bf U}_t=e^{-it{\bf H}},\ t\in {\mathbb R})$ is a strong continuous unitary group. Since $\cH$ has infinite dimension,  we should take a closure in~(\ref {uslovie}) of the linear span of ${\bf U}_t{\bf Q}_{\beta }{\bf U}_t^*$ to guarantee that  it generates a graph $\mathcal V$. In the finite dimensional case~(\ref {graph}) taking closure is not necessary. 

Following to~\cite{Weaver}, we give the following definition.

\begin{definition} An orthogonal projection ${\bf P}$ such that $\rank{\bf P}\ge 2$ is a {\bf quantum anticlique} for a non-commutative graph $\mathcal V$ if is satisfies:
\begin{equation}\label {anti}
\dim{\bf P}{\mathcal V}{\bf P}=1.
\end{equation}
\end{definition}
Using the spectral order on the set of Hermitian operators ($A>B$ iff $A-B>0$) we can give the following notion for considering a maximal error correction code.

\begin{definition} A quantum anticlique $\bf P$ for a non-commutative graph $\mathcal V$ is called {\bf maximal} if there does not exist
a quantum anticlique $\bf {\hat P}$ for $\mathcal V$ such that $\bf P<\hat P$.
\end{definition}

In the next section we consider a two-mode quantum oscillator with the Hilbert space $\cH=L^2({\mathbb R}^2)\cong L^2({\mathbb R})\otimes L^2({\mathbb R})$. We show that the inclusion~(\ref {uslovie2})  of the identity operator in  $\mathcal V$ can take place for a two-mode quantum oscillator
described by the Hamiltonian
\begin{equation}\label{Hamilt}
{\bf H}=\frac {{\bf p_1}^2}{2}+\frac {{\bf p_2}^2}{2}+\frac {({\bf q_2-q_1})^2}{2},
\end{equation}
where ${\bf q_j}$ and ${\bf p_j}$ are the position and momentum operators for ${\bf j}$th oscillator, ${\bf j}=1,2$.
The Hamiltonian (\ref {Hamilt}) is associated with the dynamics of a diatomic molecule. The corresponding operators $\bf Q$ are found and the associated operator graphs~(\ref{uslovie}) are described in detail. Maximal quantum anticliques for these operator graphs are found. For this system the infinite dimensional projection $\bf P$ is generated by one-dimensional entangled projections, while ${\bf Q}$ is a linear envelope of separable projections.

\section{Operator graphs and maximal anticliques for a pair of coupled quantum oscillators}

\subsection{Explicit action of the group ${\bf U}_t$}\label{Sec:Explicit}

We begin by deriving an explicit action of the unitary group ${\bf U}_t$. 

Equation~(\ref {sch}) in the coordinate representation has the form
\begin{equation}\label{xy}
i\frac {\partial \psi}{\partial t}=-\frac {1}{2}\frac {\partial ^2\psi }{\partial x^2}-\frac {1}{2}\frac {\partial ^2\psi }{\partial y^2}+\frac {(x-y)^2\psi }{2}.
\end{equation}
Making change of the variables $\tilde x=x+y$ and $\tilde y=x-y$
transforms this equation into
\begin{equation}\label{new}
i\frac {\partial \psi }{\partial t}=-\frac {\partial ^2\psi }{\partial \tilde x^2}-\frac {\partial ^2\psi}{\partial \tilde y^2}+\frac {\tilde y^2}{2}\psi .
\end{equation}
Here the operator
$$
-\frac {\partial ^2}{\partial \tilde y^2}+\frac {1}{2}\tilde y^2
$$
is the Hamiltonian of a one-dimensional oscillator with mass $m=1/2$ and frequency $\omega=\sqrt{2}$. Its eigenstates and the corresponding eigenvalues are
$$
\psi_n(\tilde y)=\frac {1}{\sqrt{2^nn!}(\sqrt 2\pi)^{\frac{1}{4}}}H_n\left (\frac {\tilde y}{\sqrt[4]{2}}\right )e^{-\frac{\tilde y^2}{2\sqrt 2}},\qquad \lambda_n=\sqrt {2}\left (\frac{1}{2}+n\right ),
$$
where
$$
H_n(\tilde y)=(-1)^ne^{\tilde y^2}\frac {d}{d\tilde y^n}\left (e^{-\tilde y^2}\right )$$
are Hermite polynomials, $ 
n=0,1,2,\dots $
.

By representing a solution to~(\ref{new}) in the form $\psi (t,\tilde x,\tilde y)=\sum \limits _{n=0}^{+\infty }c_n(t,\tilde x)\psi _n(\tilde y)$ we obtain
the following Cauchy problem for the coefficients $c_n(t, \tilde x)=\langle\psi_n(\cdot),\psi(t,\tilde x,\cdot)\rangle$:
\begin{eqnarray*}
i\frac{\partial c_n(t, \tilde x)}{\partial t}&=&-\frac {\partial ^2 c_n(t, \tilde x) }{\partial \tilde x^2}+\frac{2n+1}{\sqrt{2}}c_n(t, \tilde x)\\
c_n(0,\tilde x)&=&\langle\psi_n(\cdot),\psi(0,\tilde x,\cdot)\rangle
\end{eqnarray*}
Its solution is given by the Fresnel integral
$$
c_n(t,\tilde x)=\frac {1}{2\sqrt {i\pi t}}e^{-it\left(\frac{2n+1}{\sqrt{2}}\right)}\int \limits _{-\infty }^{+\infty}e^{\frac {i(\tilde x-v)^2}{4t}}c_n(0,v)dv
$$ 
Thus the solution to~(\ref {xy}) is given by the formula
\begin{eqnarray}
({\bf U}_t\psi )(x,y)&=&\sum \limits _{n=0}^{\infty}\frac {1}{2^{\frac{9}{4}}\pi \sqrt {it}2^nn!} e^{-it\left(\frac{2n+1}{\sqrt{2}}\right)}H_n\left(\frac{x-y}{2^{\frac{1}{4}}}\right)e^{-\frac {(x-y)^2}{2\sqrt{2}}}\nonumber\\
&&\times\int \limits _{-\infty }^{+\infty}\int \limits _{-\infty }^{+\infty }e^{\frac {i(x+y-(v+u))^2}{4t}}H_n\left(\frac{u-v}{2^{\frac{1}{4}}}\right)e^{-\frac {(u-v)^2}{2\sqrt{2}}}\psi (u,v)dudv. \quad \label{prop}
\end{eqnarray}

\begin{lemma} The functions $f_n(y)=\frac {1}{\sqrt {n!2^n}}H_n(y)e^{-\frac {y^2}{2}}$ 
satisfy the following equality
\begin{eqnarray}
\int\limits_{-\infty}^{+\infty} e^{-\frac {ixy}{2t}}e^{i\frac {y^2}{4t}}f_n(y)dy=\frac {\sqrt {4\pi t i}}{\pi ^{1/4}\sqrt {1+2ti}\sqrt {n!2^n}e^{i\frac {x^2}{4t}}}H_n\left (\frac {x}{\sqrt {1+2ti}}\right )e^{-\frac {x^2}{2(1+2ti)}}\label{orthog}
\end{eqnarray}
\end{lemma}
\noindent{\bf Proof.} Consider the integral operator $K:L^2({\mathbb R})\to L^2({\mathbb R})$ determined by the formula
\[
(Kf)(x)=\int \limits _{-\infty }^{+\infty}K(x,y)f(y)dy
\]
with the kernel
\[
K(x,y)=\frac {1}{\sqrt {4\pi t i}}e^{i\frac {x^2}{4t}}e^{-\frac {ixy}{2t}}e^{i\frac {y^2}{4t}}.
\]
Note that the integral operators with the kernels given below coincide
\begin{eqnarray*}
\frac {d}{dx}K(x,y)&=&\frac {ix}{2t}K(x,y)+K(x,y)\left(-\frac {iy}{2t}\right)\\
K(x,y)\frac {d}{dy}&=&\frac {ix}{2t}K(x,y)+K(x,y)\left(-\frac {iy}{2t}\right).
\end{eqnarray*}
It follows that
$$
\frac {d}{dx}K=K\frac {d}{dx}.
$$
On the other hand,
\begin{eqnarray*}
xK(x,y)&=&-2tiK(x,y)\frac {d}{dy}+K(x,y)y,\\
K(x,y)y&=&2ti\frac {d}{dx}K(x,y)+xK(x,y).
\end{eqnarray*}
Hence, for the creation operator $a^+=\frac {x-\frac {d}{dx}}{\sqrt 2}$ we get
$$
Ka^+=\frac {1}{\sqrt 2}\left (2ti\frac {d}{dx}K+xK-\frac {d}{dx}K\right )=b^+K,
$$
where
$$
b^+=\frac {x+(2ti-1)\frac {d}{dx}}{\sqrt 2}
$$
Define $f_0(x)=\frac {1}{\pi ^{1/4}}\exp(-x^2/2)$. Then
\begin{eqnarray*}
(Kf_0)(x)&=&\frac {1}{\pi ^{1/4}}\frac {1}{\sqrt {4\pi ti}}e^{i\frac {x^2}{4t}}\int\limits _{-\infty }^{+\infty} e^{-\frac {ixy}{2t}}e^{i\frac {y^2}{4t}}e^{-\frac {y^2}{2}}dy\\
&=&\frac {1}{\pi ^{1/4}}\frac {1}{\sqrt {4\pi ti}}e^{i\frac {x^2}{4t}}\int\limits _{-\infty }^{+\infty} \exp\left (-\frac {(\sqrt {2t-i}y+i\frac {x}{\sqrt {2t-i}})^2}{4t}\right )\exp\left (-\frac {x^2}{4t(2t-i)}\right)dy\\
&=&\frac {1}{\pi ^{1/4}}\frac {1}{\sqrt {4\pi ti(2t-i)}}e^{-\frac {x^2}{2(1+2ti)}}\int\limits _{-\infty }^{+\infty}\exp\left (-\frac {z^2}{4t}\right )dz=\frac {1}{\pi ^{1/4}\sqrt {1+2ti}}e^{-\frac {x^2}{2(1+2ti)}}\equiv g_0(x)
\end{eqnarray*}
Since 
$$
f_n(y)=\frac {1}{\sqrt {n!2^n}}H_n(y)e^{-\frac {y^2}{2}}=\frac {1}{\sqrt {n!}}(a^+)^n(f_0)(y)
$$
we obtain
\begin{eqnarray*}
K(f_n)(x)&=&K\left (\frac {(a^+)^n}{\sqrt {n!}}f_0\right )(x)=\frac {(b^+)^n}{\sqrt {n!}}K(f_0)(x)=\frac {(b^+)^n}{\sqrt {n!}}(g_0)(x)\\
&=&\frac {1}{\pi ^{1/4}\sqrt {1+2ti}\sqrt {n!2^n}}H_n\left (\frac {x}{\sqrt {1+2ti}}\right )e^{-\frac {x^2}{2(1+2ti)}}.
\end{eqnarray*}

$\Box$

Denote
\begin{equation}\label {eigen}
\psi _{lm}(x,y)=\frac {1}{\sqrt{2}\sqrt {\pi 2^ll!2^mm!}}H_l\left(\frac{x-y}{\sqrt[4]{2}}\right)H_m\left(\frac{x+y}{\sqrt[4]{2}}\right)e^{-\frac {(x-y)^2+(x+y)^2}{2\sqrt{2}}}.
\end{equation}

\begin{proposition} Unitary group ${\bf U}_t$ acts on $\psi_{lm}$ as
\begin{eqnarray*}
({\bf U}_t\psi _{lm} )(x,y)&=&\frac {1}{(2\pi)^{\frac{3}{4}}\sqrt {2^ll!2^mm!(1+\sqrt{2}ti)}} e^{-it\left(\frac{2l+1}{\sqrt{2}}\right)}\\
&&\times
H_l\left(\frac{x-y}{\sqrt[4]{2}}\right)e^{-\frac {(x-y)^2}{2\sqrt{2}}}H_m\left (\frac {x+y}{\sqrt[4]{2}\sqrt {1+\sqrt{2}ti}}\right )e^{-\frac {(x+y)^2}{2\sqrt{2}(1+\sqrt{2}ti)}}
\end{eqnarray*}
\end{proposition}
\noindent{\bf Proof.} We prove the Proposition by computing the integral
\begin{eqnarray*}
I=\frac{1}{\sqrt{it}}\int \limits _{-\infty }^{+\infty}\int \limits _{-\infty }^{+\infty }e^{\frac {i(x+y-(v+u))^2}{4t}}H_n\left(\frac{u-v}{\sqrt[4]{2}}\right)e^{-\frac {(u-v)^2}{2\sqrt{2}}}\\ H_l\left(\frac{u-v}{\sqrt[4]{2}}\right)H_m\left(\frac{u+v}{\sqrt[4]{2}}\right)e^{-\frac {(u-v)^2+(u+v)^2}{2\sqrt{2}}}dudv= \\
 \frac {\sqrt[4]{2}}{\sqrt{it}}\int \limits _{-\infty }^{+\infty}e^{\frac {i(x+y-z\sqrt[4]{2})^2}{4t}}e^{-\frac{z^2}{2}}H_m(z)\int \limits _{-\infty }^{+\infty }H_n\left(b\right)H_l\left(b\right)e^{-b^2}dbdz=\\
 \frac {\sqrt[4]{2}\delta_{nl}2^{n}n!\sqrt{\pi}}{\sqrt{it}}\int \limits _{-\infty }^{+\infty}e^{\frac {i(x+y-z\sqrt[4]{2})^2}{4t}} e^{-\frac{z^2}{2}} H_m(z)dz=\\
\frac {\sqrt[4]{2}\delta_{nl}2^{n}n!\sqrt{\pi}}{\sqrt{it}}e^{\frac {i(x+y)^2}{4t}}
\int \limits _{-\infty }^{+\infty}e^{\frac {-i(x+y)z\sqrt[4]{2}}{2t}} e^{\frac{i\sqrt{2}}{4t}z^2}e^{-\frac{z^2}{2}} H_m(z)dz=\\
\frac {2 \pi^{\frac{3}{4}}\sqrt{2^nn!}\delta_{nl}}{\sqrt {1+\sqrt{2}ti}\sqrt {2^mm!}}H_m\left (\frac {x+y}{\sqrt[4]{2}\sqrt {1+\sqrt{2}ti}}\right )e^{-\frac {(x+y)^2}{2\sqrt{2}(1+\sqrt{2}ti)}}
\end{eqnarray*}

$\Box$

Below we will use coherent states~\cite{gla, sud}. Given a complex number $\alpha\in\mathbb C$, the coherent state $\xi_\alpha\in L^2(\mathbb R)$ is an eigenvector of the annihilation operator with eigenvalue $\alpha$. Its explicit form is
\begin{equation}\label{coherent1}
\xi _{\alpha }(u)=\frac {1}{\pi^{\frac{1}{4}}}\exp\left (-\frac {|\alpha|^2}{2}\right )\exp\left (-\frac {u^2-2\sqrt{2}\alpha u+ \alpha^2}{2}\right ) 
\end{equation}
Consider in the tensor product of the Hilbert space $\cH=L^2({\mathbb R}^2)\cong L^2({\mathbb R})\otimes L^2({\mathbb R})$ the product of two coherent states
\begin{equation}\label{vectors}
\psi _{\alpha\beta }(x,y)=\frac {1}{\sqrt {2}}\xi _{\alpha }\left(\frac{x+y}{\sqrt[4]{2}}\right)\xi _{\beta }\left(\frac{x-y}{\sqrt[4]{2}}\right).
\end{equation}

As a result we have 
\begin{corollary}\label{corollary1}
$$
({\bf U}_t\psi _{\alpha \beta})(x,y)=\frac {e^{-\frac {it}{\sqrt{2}}}}{\sqrt 2\sqrt {1+\sqrt{2}ti}}\xi _{\alpha }\left (\frac {x+y}{\sqrt[4]{2}\sqrt {1+\sqrt{2}ti}}\right ) \xi _{e^{-i\sqrt{2}t}\beta }\left(\frac{x-y}{\sqrt[4]{2}}\right).  
$$
\end{corollary}

\noindent {\bf Proof.} Taking into account that
$$
\psi _{\alpha \beta }(x,y)=e^{-\frac {|\alpha |^2+|\beta |^2}{2}}\sum \limits _{l,m=0}^{+\infty }\frac {\alpha ^l\beta ^m}{\sqrt {l}\sqrt {m}}\psi _{lm}(x,y),
$$
where $\psi _{lm}$ are defined by (\ref {eigen}), and applying Proposition 1 we get the result.

$\Box$

\subsection{The graph generated by orbits of $({\bf U_t})$}\label{Sec:Graph}

It is straightforward to check that
\begin{equation} \label{wave_psi}
\psi _{\alpha \beta }(x,y)=\frac {1}{\sqrt {2}}\xi _{\alpha+\beta }\left (\frac {x}{\sqrt [4]{2}}\right )\xi _{\alpha -\beta }\left (\frac {y}{\sqrt [4]{2}}\right ).
\end{equation}
Given $\gamma \in \mathbb {C}$ denote
$$
\braket {x |\gamma }=\frac {1}{\sqrt [4]{2}}\xi _{\gamma }\left (\frac {x}{\sqrt [4]{2}}\right )
$$
the corresponding squeezed coherent state.
Taking an arbitrary $\beta \in {\mathbb C}$ let us define the operator
\begin{equation}\label{Q}
{\bf Q}_{\beta}=\frac {1}{\pi }\int \limits _{{\mathbb C}}\ket {\beta+\alpha,\, \alpha-\beta }\bra{\beta +\alpha,\, \alpha-\beta } d^2 \alpha
\end{equation}

\begin{proposition} The operator ${\bf Q}_{\beta }$ is an orthogonal projection.
\end{proposition}
\noindent{\bf Proof.} The operator is clearly positive. It follows from~(\ref {vectors}) that
\begin{equation}\label{action}
{\bf Q}_{\beta }\cong \frac {1}{\pi }\int \limits _{\mathbb C}\ket {\alpha }\bra {\alpha }d^2\alpha \otimes \ket {\beta }\bra {\beta }.
\end{equation}
Taking into account that 
\begin{equation}\label{squeezed}
\frac {1}{\pi}\int \limits _{\mathbb C}\ket {\alpha }\bra {\alpha }d^2\alpha={\bf I}_{L^2({\mathbb R})}
\end{equation}
we obtain that ${\bf Q}_{\beta }$ is an orthogonal projection to the subspace
\begin{equation}\label{subspace}
\cH_{\beta}=\overline {span \{\psi _{\alpha\beta },\ \alpha \in {\mathbb C}\}}.
\end{equation}
$\Box $

\begin{lemma}\label{lemmaa1} Fix $\varphi \in [0,2\pi )$, then the following inclusion holds
\begin{equation}\label{main}
{\bf I}\in \overline {span\{{\bf U}_t{\bf Q}_{re^{i\varphi }}{\bf U}_t^*,\ t\in {\mathbb R},\ r\in {\mathbb {R}_+} \}}.
\end{equation}
Moreover the right hand side of (\ref {main}) doesn't depend on a choice of $\varphi $.
\end{lemma}

\noindent{\bf Proof.} It follows from Corollary~\ref{corollary1} that
$$
({\bf U}_t \psi _{\alpha \beta})(x,y)=\frac {e^{-\frac {it}{\sqrt{2}}}}{\sqrt{2}\sqrt {1+\sqrt{2}ti}}\xi _{\alpha }\left (\frac {x-y}{\sqrt[4]{2}\sqrt {1+\sqrt{2}ti}}\right ) \xi _{e^{-i\sqrt{2}t}\beta }\left(\frac{x+y}{\sqrt[4]{2}}\right).   
$$
Taking into account (\ref {squeezed}) we obtain
for~(\ref {subspace})
$$
{\bf U}_t\cH_{\beta }=\cH_{e^{-i\sqrt{2}t}\beta}
$$
and
$$
{\bf U}_t{\bf Q}_{\beta}{\bf U}_t^*={\bf Q}_{e^{-i\sqrt{2}t}\beta}.
$$
Hence a convex hull of the set
$$
\{{\bf U}_t{\bf Q}_{re^{i\varphi }}{\bf U}_t^*,\ t\in {\mathbb R},\ r\in {\mathbb R}_+ \}
$$
contains the integral over all squeezed coherent vectors. The result follows.
$\Box $

The next statement immediately follows from Lemma~\ref{lemmaa1}. 

\begin{theorem} The operator space
$$
\mathcal{V}=\overline {span\{{\bf U}_t{\bf Q}_{re^{i\varphi}}{\bf U}_t^*,\ t\in {\mathbb R},\ r\in {\mathbb R}_+\}}=\overline {span\{{\bf Q}_{re^{i(-\sqrt{2}t+\varphi })},\ t\in {\mathbb R},\ r\in {\mathbb R}_+\}}
$$
is a non-commutative graph.
\end{theorem}

\subsection{Maximal anticlique}\label{Sec:Maximal}

Fix in $L^2({\mathbb R})$ a unit vector $g_0$ and an orthonormal basis  $\{\varphi _k\}$, $k=1,2,3,\dots$. Consider pairwise orthogonal 
unit vectors in $L^2({\mathbb R}^2)$ defined by the formula
\begin{equation}\label{vectors1}
\eta _k(x,y)=\varphi _k\left (\frac {x-y}{\sqrt 2}\right )g_0\left (\frac {x+y}{\sqrt 2}\right ),\qquad k=1,2,\dots 
\end{equation}

\begin{theorem}\label{2} The projection 
\[
{\bf P}=\sum_{k=1}^{\infty}\ket{\eta_k}\bra{\eta_k}
\]
is a maximal quantum anticlique for $\mathcal{V}.$
\end{theorem}

\noindent{\bf Proof.} Using the unitary equivalence like in~(\ref{action}) we get 
\begin{eqnarray*}
{\bf P}{\bf Q}_{e^{-i\sqrt{2}t}\beta}{\bf P}&\cong& \Bigl(\sum_{k=1}^{\infty}\ket{\varphi_k}\bra {\varphi _k}\otimes \ket {g_0}\bra{g_0}\Bigr)\Bigl({\bf I}\otimes \ket {e^{-i\sqrt{2}t}\beta }\bra {e^{-i\sqrt{2}t}\beta }\Bigr)\\
&&\times\Bigl(\sum_{k=1}^{\infty}\ket{\varphi_k}\bra {\varphi _k}\otimes \ket {g_0}\bra{g_0}\Bigr)\\
&=&|\braket{e^{-i\sqrt{2}t}\beta |g_0}|^2 ({\bf I}\otimes\ket{g_0}\bra{g_0})\cong c_{t,\beta}{\bf P}.
\end{eqnarray*}
Hence
\begin{equation}\label{PROP}
\dim{\bf P}\mathcal{V}{\bf P}=1.
\end{equation}
Suppose that there exists a quantum anticlique $\bf {\hat P}$ such that $\bf {P<\hat P}$. Then, it should have the form
$$
\bf {\hat P}\cong I\otimes \bf {\tilde P},
$$
where ${\tilde P}>\ket {g_0}\bra {g_0}$.
To satisfy (\ref {PROP}) we need
\begin{equation}\label{end}
{\bf \tilde P}\ket {e^{-i\sqrt{2}t}\beta }\bra {e^{-i\sqrt{2}t}\beta }{\bf \tilde P}=\tilde c_{t,\beta }{\bf \tilde P}
\end{equation}
for all $t$ and $\beta =re^{i\varphi}$.
It immediately follows from (\ref {end}) that ${\bf \tilde P}$ is one-dimensional, i.e. ${\bf \tilde P}=\ket {g_0}\bra {g_0}$. 
$\Box $

The non-commutative graph of Theorem~1 is generated by elementary errors which 
act on a density matrix $\rho $ of the oscillators as follows
\begin{equation}\label{elementary}
\rho \rightarrow {\bf Q}_{re^{i\varphi }}{\bf U}_t\rho {\bf U}_t^*{\bf Q}_{re^{i\varphi }}.
\end{equation}
Theorem~\ref {2} guarantees that encoding the information into pure states defined by (\ref {vectors1}) implies that these errors can be corrected. The reason is that these errors affect only the state $g_0\left (\frac {x+y}{\sqrt 2}\right )$ while images of $\varphi _k$ remain pairwise orthogonal and thus can be distinguished from each other.

\section{Conclusion} Motivated by the needs of quantum error correction, we extend the notion of operator graphs to infinite-dimensional Hilbert spaces. Our approach has allowed for the first time to construct operator graphs using the dynamical evolution in an infinite dimensional Hilbert space. As an explicit example, non-commutative operator graphs are constructed which are generated by the orbits of the unitary group given by the solution of the Schr\"odinger equation describing dynamics of a pair of oscillators. The quantum anticliques for this graph are found and their maximality is proved. 

\section{Acknowledgments}

This work was supported by the Russian Science Foundation (Project No. 17-11-01388) and performed in Steklov
Mathematical Institute of Russian Academy of Sciences.

\end{document}